\documentclass[final,numberedheadings]{aipproc}

\layoutstyle{6x9}

\usepackage{amsmath,amssymb,amsfonts,amsthm,amscd}
\usepackage{hyperref}
\usepackage[mathscr]{eucal}

\newtheorem{thm}{Theorem}
\newtheorem{prop}{Proposition}

\newtheorem*{coro}{Corollary}
\theoremstyle{definition}
\newtheorem{de}{Definition}
\newtheorem{ex}{Example}
\newtheorem{rem}{Remark}

\newcommand{\co}{\colon\thinspace}

\renewcommand{\geq}{\geqslant}

\DeclareMathOperator{\Mult}{\mathfrak{A}} \renewcommand{\O}{\Omega}

\DeclareMathOperator{\ad}{ad}

\DeclareMathOperator{\Vect}{\mathfrak{X}}

\newcommand{\der}[2]{{\frac{\partial {#1}}{\partial {#2}}}}

\newcommand{\dder}[3]{{\frac{\partial^2 {#1}}{\partial {#2}\partial {#3}}}}

\newcommand{\RR}{\mathbb R}
\newcommand{\Z}{{\mathbb Z_{2}}}
\newcommand{\ZZ}{{\mathbb Z}}

\newcommand{\p}{\partial}
\newcommand{\fun}{C^{\infty}}

\newcommand{\f}{\varphi}

\renewcommand{\a}{{\alpha}}

\renewcommand{\d}{{\delta}}

\newcommand{\la}{{\lambda}}
\newcommand{\x}{{\xi}}

\renewcommand{\t}{{\theta}}

\renewcommand{\k}{{\varkappa}}

\newcommand{\e}{{\varepsilon}}

\renewcommand{\o}{{\omega}}

\newcommand{\s}{{\sigma}}

\newcommand{\kt}{{\tilde k}}

\newcommand{\lt}{{\tilde l}}
\newcommand{\at}{{\tilde a}}
\newcommand{\bt}{{\tilde b}}
\newcommand{\ct}{{\tilde c}}

\newcommand{\ft}{{\tilde f}}

\newcommand{\Pt}{{\tilde P}}

\newcommand{\pt}{{\tilde p}}
\newcommand{\qt}{{\tilde q}}

\begin{document}

\title{Higher Poisson Brackets and Differential Forms}

\classification{02., 02.20.Tw, 02.40.-k, 45.20.Jj} \keywords {Higher
Poisson bracket, strongly homotopy Lie algebra, supermanifold,
symplectic form, Legendre transformation, higher Koszul bracket}

\author{H.~M.~Khudaverdian}{
  address={School of Mathematics, The University of Manchester, Oxford Road, Manchester, M13 9PL, UK} }

\author{Th.~Th.~Voronov}{
  address={School of Mathematics, The University of Manchester, Oxford Road, Manchester, M13 9PL, UK}
}

\begin{abstract}
We show how the relation between Poisson brackets and symplectic
forms can be extended to the case of  inhomogeneous multivector
fields and  inhomogeneous differential forms (or pseudodifferential
forms). In particular we arrive at a notion which is a
generalization of a symplectic structure and gives rise to higher
Poisson brackets. We also obtain a construction of  Koszul type
brackets in this setting.
\end{abstract}

\maketitle


\section{Introduction}

Consider a Poisson manifold $M$ with a Poisson tensor $P=(P^{ab})$.
It is a known fact that  raising indices with the help of
$P^{ab}$ gives the following commutative diagram:
\begin{equation}\label{intro.eq.diagram}
    \begin{CD} \Mult^k (M)@>{d_P}>> \Mult^{k+1}(M)\\
                @AAA         @AAA\\
                \O^k(M)@>{d}>> \O^{k+1}(M) \,.
    \end{CD}
\end{equation}
Here we denote  by $\Mult^k(M)$ the space of multivector fields on
$M$ of degree $k$. In the sequel we also use the notations such as
$\Mult(M)$ and $\O(M)$ for the algebras of multivector fields and
differential forms, respectively. (On supermanifolds one should
speak of `pseudodifferential forms' and `pseudomultivector fields',
but we shall stick to a simplified usage unless it may lead a
confusion.) The vertical arrows are the operations of raising
indices with the help of $P$ and the top horizontal arrow is the
Lichnerowicz differential $d_P=[P, \ \ ]$. The bracket is the
canonical Schouten bracket of multivector fields. This diagram leads
to a natural map $H^k(M, \RR)\to H^k(\Mult(M), d_P)$ from the de
Rham to Poisson cohomology, which is an isomorphism when the bracket
is symplectic.

The transformation $\O(M)\to \Mult(M)$   also preserves the
brackets, so it is a morphism  of differential Lie superalgebras.
Here the space of multivector fields $\Mult(M)$ is considered with
the canonical Schouten bracket and the space of forms $\O(M)$, with
an odd bracket known as the Koszul bracket, induced by the Poisson
structure on $M$. It is noteworthy that the differential on $\O(M)$
is canonical while the bracket depends on the Poisson tensor $P$,
and the case of $\Mult(M)$ is the opposite: the  bracket is
canonical but  the differential depends on  $P$; the map $\O(M)\to
\Mult(M)$ exchanges a canonical structure with one defined by $P$.

In this paper we   show three things. Firstly, we shall show how the
map $\O(M)\to \Mult(M)$ and the diagram~\eqref{intro.eq.diagram} can
be generalized to the case when a bivector field $P$ is replaced by
an arbitrary even multivector field, on a supermanifold. Secondly,
we shall explain what plays the role of a symplectic structure  in such  case (when a bivector $P$ is replaced by an inhomogeneous object). We shall
show how an   inhomogeneous even form with an appropriate
non-degeneracy condition generates a sequence of `higher' Poisson
brackets making the space, $\fun(M)$, a homotopy Poisson algebra.
This might be called a \textit{generalized} or \textit{homotopy
symplectic structure} on $M$. It is remarkable that the role of the
matrix inverse (for bivectors and $2$-forms) is taken by the
Legendre transform. Thirdly, we shall also explain what is the
replacement of the Koszul bracket for higher Poisson structure.

Here and in the main text, by a \textit{homotopy Poisson algebra} we
mean an $L_{\infty}$-algebra of Lada and
Stasheff~\cite{stasheff:shla93}\,---\,the superized
version\,---\,endowed with a commutative associative multiplication
w.r.t. which each bracket is a multiderivation. This is more
restrictive than other notions discussed in the
literature~\cite{tamarkintsygan:bv}, but seems quite fitting for
differential-geometric purposes. Similarly one defines a
\textit{homotopy Schouten algebra}.

The constructions that we discuss have direct analogs for an odd
Poisson structure on $M$, as well as for Lie algebroids. (In fact,
the roots of this work are in our studies of odd Laplacians
in~\cite{tv:laplace2}.) There is also a remarkable analogy with
well-known constructions of classical mechanics. A more detailed
text containing proofs will appear elsewhere.

A note about usage: to simplify the language, we usually speak about
`manifolds', `Lie algebras', etc., meaning `supermanifolds' and
`superalgebras' respectively, unless this may cause a confusion or
we need to emphasize that we are dealing with a `super' object. By a
\textit{$Q$-manifold} we mean a \textit{differential manifold},
i.e., a supermanifold with a homological vector field. The reader
should be warned that parity of objects (i.e., $\Z$-grading) and a
$\ZZ$-grading such as degree of forms, where it make sense, are in
general independent; when we speak about an object which is `even',
that means `even in the parity sense'. We use different notations
for different types of brackets; the canonical Schouten brackets and
the Koszul-type brackets of forms  are denoted by the square
brackets, while the canonical Poisson brackets, by the parentheses
(round brackets). All other Poisson-type brackets are denoted by the
braces (curly brackets). A subscript may be used to indicate a
Poisson-type structure defining the bracket.

\section{Main constructions}

Let $M$ be a manifold (or supermanifold, which we shall still call a
manifold according to our convention). There are two supermanifolds
naturally associated with it: the tangent bundle with the reversed
parity $\Pi TM$ and the cotangent bundle with the reversed parity
$\Pi T^*M$. We denote by $\Pi$ the parity reversion functor.

Recall that for an ordinary manifold $M$, the differential forms on
$M$ can be identified with the functions on the supermanifold $\Pi
TM$, and the multivector fields on $M$, with the functions on $\Pi
T^*M$,
\begin{align}
    \O(M)&=\fun(\Pi TM)\,,\\
    \Mult(M)&=\fun(\Pi T^*M)\,.
\end{align}
For a supermanifold $M$ we shall take these as the definitions. (This
simple approach will be sufficient for our needs. We are not going
into a deeper investigation of  analogs of differential forms on
supermanifolds here.)

If $x^a$ are local coordinates on $M$, then $dx^a$ and $x^*_a$ are
the induced coordinates in  the fibers of the vector bundles $\Pi
TM$ and $\Pi T^*M$ respectively. They have parities opposite to that
of $x^a$: $\widetilde{dx^a}=\widetilde{x^*_a}=\at+1$. (We use the
tilde to denote parity and $\at$ stands for the parity of the
coordinate $x^a$.) The transformation laws for them are
\begin{equation*}
    dx^a =dx^{a'}\,\der{x^a}{x^{a'}} \quad \text{and} \quad
     x^*_a =\der{x^{a'}}{x^{a}}\,x^*_{a'}
\end{equation*}
(mind the order).

Let us fix an arbitrary even multivector field $P\in \Mult(M)$. We
shall define a bundle map
\begin{equation*}
    \f_{P}\co \Pi T^*M\to \Pi TM
\end{equation*}
by the formula
\begin{equation}\label{main.eq.fip}
    \f_{P}^*(dx^a)=(-1)^{\at+1}\der{P}{x^*_a}\,.
\end{equation}
One can show that the map is well-defined.
\begin{ex} If $P=\frac{1}{2}\,P^{ab}(x)x^*_bx^*_a$, then the
pull-back $\f_{P}^*\co \O(M)\to \Mult(M)$ coincides with  raising
indices with the help of the tensor $P^{ab}$.
\end{ex}

Now we shall study an analog of the
diagram~\eqref{intro.eq.diagram}. Consider the diagram
\begin{equation}\label{main.eq.diagram}
    \begin{CD} \Mult (M)@>{d_P}>> \Mult(M)\\
                @A{\f_P^*}AA         @AA{\f_P^*}A\\
                \O(M)@>{d}>> \O(M) \,,
    \end{CD}
\end{equation}
where $d_P:=\ad P=[P,\ \ ]$ (the Schouten bracket). The linear
operator $d_P$ is odd. In general $d_P^2\neq 0$ and
\begin{equation*}
    d_P^2=\frac{1}{2}\,\ad [P,P]\,.
\end{equation*}
The diagram~\eqref{main.eq.diagram} is,  in general, not
commutative. To describe its discrepancy we need one technical tool.

Any multivector field $Q\in \Mult(M)=\fun(\Pi T^*M)$ defines a
derivation from the algebra $\O(M)$ to the tensor product
$\O(M)\otimes_{\fun(M)} \Mult(M)$ over the natural homomorphism
$\O(M)\to \O(M)\otimes_{\fun(M)} \Mult(M)$, $\o\mapsto \o\otimes 1$.
We denote it $\varkappa_Q$. In coordinates,
\begin{equation}\label{main.eq.kappaq}
    \varkappa_Q=(-1)^{(\tilde Q+1)(\tilde a+1)} \der{Q}{x^*_a}\,\der{}{dx^a}\,.
\end{equation}

\begin{thm} \label{main.thm.discr} The discrepancy of the diagram~\eqref{main.eq.diagram}
is given by the formula:
\begin{equation}\label{main.eq.discrep}
    \f_P^*\circ d- d_P\circ \f_P^*=-\frac{1}{2}\,\f_P^*\k_{[P,P]}\,.
\end{equation}
\end{thm}
\begin{coro} If $[P,P]=0$, then the diagram~\eqref{main.eq.diagram}
is commutative and thus $\f_P$ is a map of $Q$-manifolds
\begin{equation*}
    (\Pi T^*M, d_P)\to (\Pi TM, d)\,.
\end{equation*}
\end{coro}

An even multivector field $P$ satisfying $[P,P]=0$ is a
generalization of a Poisson tensor. It defines a sequence of `higher
Poisson brackets' on $M$, i.e., a sequence of $n$-ary operations,
$n=0,1, 2, 3\ldots \ $, on the space $\fun(M)$,
\begin{equation}\label{main.eq.higherpoisson}
    \{f_1,\ldots,f_n\}_P:=[\ldots [P,f_1],\ldots,f_n]|_M\,.
\end{equation}
(Although it is not manifest in the formula, the bracket is
antisymmetric in $f_1,\ldots,f_n$.) When no  confusion is possible
we suppress the subscript $P$ for the Poisson brackets. Each
operation is a multiderivation w.r.t. the associative
multiplication, i.e., a derivation in each argument. The condition
$[P,P]=0$ ensures that $\fun(M)$ becomes an $L_{\infty}$-algebra
w.r.t. the brackets. (See~\cite{tv:laplace2}, \cite{tv:higherder}.)

For an ordinary Poisson structure on a manifold $M$, it is
non-degenerate if the components of the Poisson bivector $P^{ab}$
make a non-degenerate matrix. Then the entries of the inverse matrix
$(P^{ab})^{-1}$ are the components of a non-degenerate closed
$2$-form. Conversely, any non-degenerate closed $2$-form $\o$
defines a Poisson bracket for which the components of the Poisson
bivector are the entries of the inverse matrix $(\o_{ab})^{-1}$.
This is the relation between Poisson brackets and symplectic
structures. How one should extend it to the case of inhomogeneous
(and, possibly, even not fiberwise-polynomial, in the super case)
multivector fields and forms?

\begin{prop} The map $\f_{P}\co \Pi T^*M\to \Pi TM$ is a
diffeomorphism (at least, near the zero section) if the matrix of
second partial derivatives
\begin{equation*}
    \dder{P}{x^*_a}{x^*_b}
\end{equation*}
is invertible at $x^*_a=0$.
\end{prop}

This is what replaces the non-degeneracy condition for a bivector
field. What should stand for the inverse matrix?

Suppose the map $\f_{P}\co \Pi T^*M\to \Pi TM$ is invertible.
Consider the  \textbf{fiberwise Legendre transformation}  of the multivector
field $P$:
\begin{equation}\label{main.eq.legp}
    \o:= \check{{P}}:=(\f_P^*)^{-1}\left((-1)^{\at+1}\der{P}{x^*_a}x^*_a-P(x,x^*)\right)\,
\end{equation}
(the first term in brackets is just $dx^a x^*_a$ if we apply the corresponding isomorphism).  It is an even differential form on $M$. By changing the order in the first term,
\begin{equation}\label{main.eq.legp2}
    \o= \check{{P}}=(\f_P^*)^{-1}\left(x^*_a\der{P}{x^*_a}-P(x,x^*)\right)\,.
\end{equation}

\begin{prop}
The inverse map $\f_{P}^{-1}\co \Pi TM\to \Pi T^*M$ is defined by
the form $\o$ in the similar way as the original map is defined by
$P$:
\begin{equation*}
    \f_{P}^{-1}=\psi_{\o} \quad \text{where} \quad
    \psi_{\o}^*(x^*_a)=\der{\o}{dx^a}\,.
\end{equation*}
The multivector field $P$ can be recovered as the fiberwise Legendre transformation of the form $\o$.
\end{prop}

This statement follows from the well known properties of the
Legendre transformation. (Geometrically, we have a Lagrangian
submanifold in each fiber of the sum $\Pi TM\oplus \Pi T^*M$, which
can be  described as the graph of the `gradient' of a function of
either $x^*_a$ or $dx^a$; then the corresponding functions are
related by the mutually-inverse Legendre transforms.)

\begin{ex} For the classical case $P$ is quadratic,
    $P=\frac{1}{2}\,P^{ab}x^*_bx^*_a\,$,
therefore
    $x^*_a\der{P}{x^*_a} -P = P$
and the Legendre transformation of $P$ is simply $(\f_P^*)^{-1}(P)$.
We have
\begin{equation*}
    \f_P^*(dx^a)=(-1)^{\at+1}\der{P}{x^*_a}=P^{ab}x^*_b\,, \quad \text{so} \quad (\f_P^*)^{-1}x^*_a=P_{ab}dx^b\,,
\end{equation*}
where $P_{ab}$ stand for the matrix entries of the inverse matrix for $P^{ab}$. Hence we arrive at the $2$-form
\begin{equation*}
    \o=  \frac{1}{2}\,dx^aP_{ab}dx^b =\frac{1}{2}\,dx^adx^b\,P_{ab}(-1)^{\at(\bt+1)}=\frac{1}{2}\,dx^adx^b\,P_{ba}(-1)^{\at +1}
\end{equation*}
or
\begin{equation*}
    \o=  \frac{1}{2}\,dx^adx^b\,\o_{ba} \quad \text{where} \quad \o_{ab}=P_{ab}(-1)^{\bt+1}\,.
\end{equation*}
(Note the symmetry properties: $P^{ab}=(-1)^{(\at+1)(\bt+1)}P^{ba}$,
$\o_{ab}=(-1)^{(\at+1)(\bt+1)}\o_{ba}$, and
$P_{ab}=-(-1)^{\at\bt}P_{ba}$.)
\end{ex}

The following theorem is an extension of the classical relation existing for $2$-forms and bivector fields.

\begin{thm} The exterior differential of the  Legendre
transform  of the multivector field $P$, the form  $\o$,  is given by the
formula:
\begin{equation}\label{main.eq.domega}
    d\o=-\frac{1}{2}(\f_P^*)^{-1}\bigl([P,P]\bigr)\,.
\end{equation}
\end{thm}
\begin{coro} The multivector $P$ satisfies $[P,P]=0$ if and only if
$d\o=0$.
\end{coro}

\begin{rem} The  statement in one direction can be also deduced  from Theorem~\ref{main.thm.discr}.
Suppose a form $\o'$ maps to the multivector field $P$ under the map
$\f_P^*$. Under the assumption that $[P,P]=0$, the diagram~\eqref{main.eq.diagram} commutes
and $\f_P^*(d\o')=d_P(\f^*_P\o')=d_P(P)=0$. Assuming that $\f_P^*$
is invertible, we arrive at $d\o'=0$. The question arises, what is
the relation between the form $\o'=(\f_P^*)^{-1}P$ and the form
$\o=\check P$, the Legendre transform of $P$. The formula~\eqref{main.eq.legp2} for the
Legendre transform can be re-written as follows:
\begin{equation}\label{main.eq.legpeuler}
    \o=  (\f_P^*)^{-1}\bigl(E(P)-P\bigr)\,.
\end{equation}
where $E$ is the fiberwise Euler vector field on $\Pi T^*M$. In the classical case $\o$ and $\o'$ coincide, as we have seen, since  for a bivector $E(P)=2P$.  We may note the following useful relation for arbitrary multivector fields:
\begin{equation}
    E\bigl([P,Q]\bigr)=[E(P),Q]+[P,E(Q)]-[P,Q]
\end{equation}
(which is an expression of the fact that the Schouten bracket has weight $-1$ in the natural $\ZZ$-grading). Therefore  $[P,E(P)]=\frac{1}{2}\,([E(P),P]+[P,E(P)])=\frac{1}{2}\,(E[P,P]+[P,P])$ and
    $[P,E(P)-P]=\frac{1}{2}\,(E[P,P]-[P,P])\,$.
We see that if $[P,P]=0$, then $[P,E(P)]=0$ too and the forms $\o$
and $\o'$ are both closed. The choice of $\o$, not $\o'$, as the
correct analog of a symplectic form corresponding to $P$ is
determined by the fact that the inverse map $(\f_P)^{-1}$ is given
by $\psi_{\o}$ defined from $\o$ in the same way as $\f_P$ is
defined from $P$, with the same type of non-degeneracy conditions.
\end{rem}

The above constructions can be summarized in the following definitions.

\begin{de} A higher Poisson structure on $M$ given by a multivector field $P$ is \textit{non-degenerate}
if the map $\f_P\co \Pi T^*M\to \Pi TM$ is invertible (at least on a
neighborhood of $M$ in $\Pi T^*M$).
\end{de}

In terms of the higher Poisson brackets $\{f_1,\ldots,f_k\}_P$ generated
by $P$, this is  equivalent to the non-degeneracy of just the binary
bracket $\{f,g\}_P$. Since however this bracket  satisfies the Jacobi
identity only up to homotopy, it is not the same as an ordinary
symplectic structure.

\begin{de} A \textit{generalized} or \textit{homotopy, symplectic structure} on $M$ is a closed even (pseudo)differential form $\o\in \O(M)$ such that the map $\psi_{\o}\co \Pi TM\to \Pi T^*M$ defined by the formula
\begin{equation}\label{main.eq.psiomega}
    \psi_{\o}^*x^*_a=\der{\o}{dx^a}
\end{equation}
is a diffeomorphism.  (As above, we   relax this condition by
requiring  $\psi_{\o}$ to be a diffeomorphism only on a
neighborhood of the zero section $M\subset \Pi TM$.) Such a form
$\o$ is called a \textit{generalized symplectic form}.
\end{de}

The generalized symplectic forms   $\o\in \O(M)$ are in one-to-one
correspondence with the non-degenerate higher Poisson structures given
by multivector fields $P\in \Mult(M)$ and the correspondence is
given by the mutually-inverse Legendre transforms.

\begin{prop} The non-degeneracy condition for a generalized symplectic form is equivalent to requiring that the matrix of
second  partial derivatives
\begin{equation*}
    \dder{\o}{dx^a}{dx^b}
\end{equation*}
is invertible at $dx^a=0$.
\end{prop}

\begin{ex}
If an even form $\o\in \O(M)$ can be written as
\begin{equation}\label{main.eq.inhomform}
    \o=\o_0+\o_1+\o_2+\o_3+\ldots
\end{equation}
where $\o_k\in\O^k(M)$, it is a generalized symplectic form if and
only if $\o_2$ is an ordinary symplectic form and the other terms
are arbitrary closed forms. (Note that on a supermanifold an even in
the sense of parity form may have components  both in even and odd
degrees.) As we shall see in the next section, the Poisson brackets
defined by~\eqref{main.eq.inhomform} will not reduce to the ordinary
Poisson bracket defined by the symplectic $2$-form $\o_2$, but
include `higher corrections'.
\end{ex}

\section{Examples}

Let $\o\in\O(M)$ be a generalized symplectic form on $M$ with the corresponding non-degenerate Poisson multivector field $P\in\Mult(M)$. What are the (higher) Poisson brackets generated by this form?

For an arbitrary higher Poisson structure defined by $P\in \Mult(M)$, not necessarily non-degenerate, the higher Poisson brackets of functions on $M$ are given by~\eqref{main.eq.higherpoisson}. The Hamiltonian vector fields of ordinary Poisson geometry are replaced by   multivector fields.  For a function $f\in \fun(M)$, the multivector field $Q_f\in \Mult(M)$ defined by the formula
\begin{equation}\label{exa.eq.hamfield1}
    Q_f:=\f_P^*(df)
\end{equation}
may be called the \textit{Hamiltonian multivector field} corresponding to $f$. By Theorem~\ref{main.thm.discr},
\begin{equation}\label{exa.eq.hamfield2}
    Q_f=[P,f]\,.
\end{equation}
Indeed, from the commutative diagram, $Q_f=\f_P^*(df)= d_Pf=[P,f]$, because for functions $\f_P^*(f)=f$. The formula~\eqref{main.eq.higherpoisson} for higher Poisson brackets may be reformulated as
\begin{equation}\label{exa.eq.brackets}
    \{f_1,\ldots,f_k\}=[\ldots [[Q_{f_1},f_2],\ldots,f_k]|_M\,.
\end{equation}
Let $\o$ be a generalized symplectic form. To calculate the Poisson
brackets corresponding to  $\o$, it is sufficient to find the
multivector field $Q_f$ for an arbitrary function $f\in\fun(M)$. We
find it from the relation
\begin{equation*}
    Q_f=(\psi_{\o}^*)^{-1}(df) \quad \text{or} \quad \psi_{\o}^*Q_f=df\,.
\end{equation*}

Let us consider particular examples.

\begin{rem}
Although the case of ordinary manifolds is not at all
trivial, more interesting examples should be related with
supermanifolds. Indeed, for an ordinary manifold, $\o$ is just an
inhomogeneous differential form of the
appearance~\eqref{main.eq.inhomform}. Since $\o$ is supposed to be
even, then only the $\o_{2ks}$ may be non-zero. As we shall see, this
implies the vanishing of the differential, i.e., the unary bracket,
on functions. However the higher brackets may still be non-zero and
satisfy non-trivial identities.
\end{rem}

\begin{ex} We start from an ordinary symplectic structure for further comparison.
Suppose
\begin{equation*}
    \o=\o_2=\frac{1}{2}\,dx^adx^b\o_{ba}(x)\,.
\end{equation*}
Then we have the equation
\begin{equation*}
    x^*_a=dx^b\o_{ba}
\end{equation*}
for determining the variables $dx^a$. Here and in the sequel we
shall suppress the notations for the pull-backs   $\psi_{\o}^*$ and
$\f_P^*$. From here
\begin{equation*}
    dx^a=x^*_b\o^{ba}=(-1)^{\at +1}\o^{ab}x^*_b
\end{equation*}
where $\o_{ac}\o^{cb}=\d_a^b$ (note that
$\o^{ab}=-(-1)^{\a\bt}\o^{ba}$), and
\begin{equation*}
    Q_f=x^*_b\o^{ba}\p_a{f} =(-1)^{\ft(\at+1)}\p_a{f}x^*_b\o^{ba}=-(-1)^{(\at+1)\ft}\p_a{f}\o^{ab}x^*_b\,.
\end{equation*}
Therefore the only non-trivial bracket is, of course, binary, and it is given by
\begin{equation*}
    \{f,g\}=-(-1)^{\ft(\at+1)}\p_a{f}\o^{ab}\p_b{g} =(-1)^{\ft\at+1}\o^{ab}\p_b{f}\p_a{g}\,.
\end{equation*}
The Poisson tensor is given by
\begin{equation*}
    P=\frac{1}{2}\,(-1)^{\at +1}\o^{ab}x^*_bx^*_a\,.
\end{equation*}
\end{ex}

\begin{ex} Suppose now there is an extra linear term in $\o$:
\begin{equation*}
    \o=\o_1+\o_2=dx^a\o_a+\frac{1}{2}\,dx^adx^b\o_{ba}\,.
\end{equation*}
Note that in particular $d\o_1=0$, hence locally $\o_1=d\chi$ for some odd function $\chi$. We have the equation
\begin{equation*}
    x^*_a=\o_a+dx^b\o_{ba}
\end{equation*}
for determining   $dx^a$ (where locally $\o_a=\p_a\chi$). As before
we obtain
\begin{equation*}
    dx^a=(x^*_b-\o_b)\o^{ba}=(x^*_b-\p_b\chi)\o^{ba}\,.
\end{equation*}
It is instructive to find the Poisson multivector field $P$. When we
calculate the Legendre transform, the term $\o_1$ makes no input
into $E(\o)-\o$ and  results only in the shift of the argument:
\begin{equation*}
    P=\frac{1}{2}\,(-1)^{\at +1}\o^{ab}(x^*_b-\p_b\chi)(x^*_a-\p_a\chi)\,.
\end{equation*}
Hence we have
\begin{equation*}
    P=P_0+P_1+P_2=\frac{1}{2}\,(-1)^{\at +1}\o^{ab}\p_b\chi\p_a\chi -\,(-1)^{\at +1} \o^{ab}\p_b\chi\, x^*_a+\frac{1}{2}\,(-1)^{\at +1}\o^{ab}x^*_bx^*_a\,.
\end{equation*}
This leads to the following $0$-, $1$-, and $2$-brackets:
\begin{align*}
    \{\varnothing\}&=\frac{1}{2}\,(-1)^{\at +1}\o^{ab}\p_b\chi\p_a\chi=\frac{1}{2}\,\{\chi,\chi\}\,,\\
    \{f\}&=(-1)^{\at +1} \o^{ab}\p_b\chi\p_a{f}=\{\chi,f\}\,,\\
    \{f,g\}&=(-1)^{\ft\at+1} \o^{ab}\p_b{f}\p_a{g}\,,
\end{align*}
and there are no higher brackets. Hence we have  the binary
Poisson bracket that satisfies the ordinary Jacobi identity. Besides
it we are given an odd vector field $X=X_{\chi}$ locally-Hamiltonian
w.r.t. this bracket (and thus automatically a derivation)  and an
even function $P_0=\frac{1}{2}\,\{\chi,\chi\}$ such that
$X^2=X_{P_0}$. The field $X$ is  homological  if $\{\chi,\chi\}$ is
a local constant. (One may consider, unrelatedly to generalized symplectic structures, a  structure similar
to the above consisting of a Poisson bracket together with an odd function $\chi$ defining  $0$- and $1$-brackets by the above formulas, where the corresponding vector field is homological  if $\{\chi,\chi\}$ is a Casimir function.)
\end{ex}

\begin{ex} Consider now a generalized symplectic form
\begin{equation*}
    \o=\o_1+\o_2+\o_3=dx^a\o_a+\frac{1}{2}\,dx^adx^b\o_{ba}(x)+\frac{\la}{3!}\,dx^adx^bdx^c\o_{cba}\,
\end{equation*}
involving a cubic term. Notice that we have included a parameter $\la$.
We may start as before and obtain the relation
\begin{equation*}
    x^*_a=\o_a+dx^b\o_{ba}+\frac{\la}{2}dx^bdx^c\o_{cba}
\end{equation*}
for determining   $dx^a$. In order to obtain the solution introduce
$\x_a=dx^b\o_{ba}$. Hence $dx^a=(-1)^{\at +1}\o^{ab}\x_b$, and we
have the equation
\begin{equation*}
    \x_a+\frac{\la}{2}\,\o^{kl}_a\x_l\x_k=\t_a
\end{equation*}
for determining $\x_a$ (if we denote $\t_a=x^*_a-\o_a$). Here we
raise indices with the help of $\o^{ab}$ with the following sign
convention:
\begin{equation*}
    \o^{kl}_a=\o^{kb}\o^{lc}\o_{cba}(-1)^{(\lt+1)(\bt+1)+\at(\kt+\lt)}\,.
\end{equation*}
The equation for  $\x_a$ can be solved by iterations,  expressing the answer as an
infinite power series in $\la$. In the first order in $\la$,
\begin{equation*}
    dx^a=(-1)^{\at +1}\o^{ab}(x^*_b-\o_b)-\frac{\la}{2}(-1)^{\at
    +\ct}\o^{abc}(x^*_c-\o_c)(x^*_b-\o_b)+O(\la^2)\,.
\end{equation*}
Here
\begin{equation*}
    \o^{abc}=\o^{ap}\o^{bq}\o^{cr}\o_{rqp}(-1)^{\pt(\bt+\ct)+\qt(\ct+1)}\,.
\end{equation*}
To obtain the Legendre transform, this should be substituted into
$E(\o)-\o=\o_2+2\o_3$. We arrive at
\begin{equation*}
    P= P_0+P_1+P_2+P_3+\ldots
\end{equation*}
where
\begin{align*}
    P_0&=\frac{1}{2}\,(-1)^{\at+1}\o^{ab}\o_b\o_a+\frac{\la}{6}\,(-1)^{\at+\ct}\o^{abc}\o_c\o_b\o_a+O(\la^2)\,,\\
    P_1&=\left((-1)^{\at}\o^{ab}\o_b-\frac{\la}{2}\,(-1)^{\at+\ct}\o^{abc}\o_c\o_b+O(\la^2)\right)x^*_a\,,\\
    P_2&=\frac{1}{2}\,\left((-1)^{\at+1}\o^{ab}+ {\la} \,(-1)^{\at+\ct}\o^{abc}\o_c+O(\la^2)\right)x^*_bx^*_a\,,\\
    P_3&=-\frac{\la}{6}\,(-1)^{\at+\ct}\o^{abc}x^*_cx^*_bx^*_a+O(\la^2)\,, \\
    P_4&=O(\la^2)\,, \ \text{etc.}
\end{align*}
Therefore there will be an infinite series of brackets and each
bracket is given by an infinite series in the parameter $\la$; all
brackets higher than ternary are of order $\geq 2$ in   $\la$. They
satisfy non-trivial Jacobi identities with $n$ arguments for all
$n=0,1,2,3, \ldots$ \ . Note the presence of higher corrections in
the binary bracket.
\end{ex}

\section{Higher Koszul brackets}

We shall discuss now what replaces the Koszul bracket in the case of
a higher Poisson structure, i.e., an even multivector field $P\in
\Mult(M)$ such that $[P,P]=0$.

Let us recall the ordinary case. In the classical situation the
Koszul bracket corresponding to a Poisson structure on a manifold
$M$ given by a bivector field $P$ may be defined axiomatically as a
unique odd Poisson (Schouten, Gerstenhaber, ...) bracket on the
algebra of forms $\O(M)$ obeying the following `initial conditions':
\begin{equation}\label{koszul.eq.binaryabstr}
    [f,g]_P=0\,, \quad [f,dg]_P=(-1)^{\ft}\{f,g\}_P\,, \quad \text{and} \quad [df,dg]_P=-(-1)^{\ft}d\{f,g\}_P\,,
\end{equation}
where the curly bracket $\{\ ,\ \}_P$ stands for the Poisson bracket
of functions and $[\ ,\ ]_P$ stands  for the Koszul bracket of
forms.  In particular, for coordinates and their differentials we
have
\begin{equation}\label{koszul.eq.binary}
    [x^a,x^b]_P=0\,, \quad [x^a,dx^b]_P=-P^{ab}\,, \quad \text{and} \quad
    [dx^a,dx^b]_P= dP^{ab}\,.
\end{equation}

(The Lie bracket of $1$-forms on a Poisson manifold was probably
first introduced by B.~Fuchssteiner~\cite{fuchssteiner82}, but it
had a rich pre-history, see~\cite{yvette:modular2008}. The bracket
on the algebra of all forms was introduced by
Koszul~\cite{koszul:crochet85}, as the bracket generated by a
second-order operator on forms playing the role of the boundary
operator for the Poisson homology~\footnote{It mimics the well-known
expression of the canonical Schouten bracket on multivector fields
in terms of a divergence operator, which is also a model for the
Batalin--Vilkovisky formalism.}. See also~\cite{karasev:maslov1991}
and references therein. Particular signs in formulas such
as~\eqref{koszul.eq.binaryabstr}, \eqref{koszul.eq.binary} depend on
conventions.)

In this case the Koszul bracket can be also defined using the
diagrams~\eqref{intro.eq.diagram} or \eqref{main.eq.diagram}.
Namely, if we assume the invertibility of the matrix $(P^{ab})$,
then one can consider
\begin{equation}\label{koszul.eq.binarynaive}
    [\o,\s]_P:=(\f_P^*)^{-1}\bigl([\f_P^*\o, \f_P^*\s]\bigr)\,.
\end{equation}
By the construction it is   an odd Poisson bracket on the algebra
$\O(M)$. One can see that an explicit formula obtained
from~\eqref{koszul.eq.binarynaive} does not include the inverse
matrix for $(P^{ab})$ (this substantially relies on the identity
$[P,P]=0$) and for coordinates gives
exactly~\eqref{koszul.eq.binary}. Therefore
formula~\eqref{koszul.eq.binarynaive} `survives the limit' when one
passes to an arbitrary Poisson bivector.

All the above holds true for the classical (i.e., binary) case only.
It is not a priori obvious how one can extend the axiomatic
definition to the general case since there are now many `higher'
Poisson brackets of functions. If one would try  to use naively the
formula~\eqref{koszul.eq.binarynaive} defining an odd binary bracket
on forms as an operation isomorphic to the canonical Schouten
bracket of multivector fields, then it would not survive the limit
when the condition of the non-degeneracy is dropped unlike for the
classical case.

Therefore we need a different approach.

One should expect that to a higher Poisson structure on functions
there corresponds a higher structure on forms as well  rather than a
single bracket, i.e., a sequence of `higher Koszul brackets'. We
shall indeed define them\,---\,directly in terms of the multivector
field $P\in \Mult(M)$.

Recall that  odd brackets on functions on a given manifold $N$ are
generated by an odd  `master Hamiltonian' $S$, i.e., a function on
the cotangent bundle $T^*N$  satisfying $(S,S)=0$ for the canonical
Poisson bracket. See, e.g.,~\cite{tv:graded}. In our case we need
odd brackets on the algebra $\O(M)=\fun(\Pi TM)$. Therefore we
should look for an \textbf{odd} function on the cotangent bundle
$T^*(\Pi TM)$. How  can one get it from a given even function $P\in
\fun(\Pi T^*M)$?

\begin{thm}
\label{koszul.thm.constr}
There is a natural odd linear map
\begin{equation}\textrm{\label{koszul.eq.map}}
   \alpha \co \fun(\Pi T^*M)\to \fun(T^*(\Pi TM))
\end{equation}
that takes the canonical Schouten bracket on $\Pi T^*M$
to the canonical Poisson bracket on $T^*(\Pi TM)$, up to a sign:
\begin{equation}\label{koszul.eq.mapofbrackets}
   \a\left([P,Q]\right)= (-1)^{\Pt+1}\bigl(\a(P), \a(Q)\bigr)\,,
\end{equation}
for arbitrary $P,Q\in \fun(\Pi T^*M)$.
\end{thm}

(Here by the square brackets we denote the canonical Schouten
bracket and by the parentheses, the canonical Poisson bracket. The sign in~\eqref{koszul.eq.mapofbrackets} depends on conventions.)

Sketch of a proof. By the theorem of Mackenzie and
Xu~\cite{mackenzie:bialg} (see also~\cite{tv:graded}) there is a
symplectomorphism between $T^*(\Pi TM)$ and $T^*(\Pi T^*M)$. Now,
given a function on $\Pi T^*M$, one can associate with it the
corresponding Hamiltonian vector field of the opposite parity w.r.t.
the canonical bracket on $\Pi T^*M$. The Schouten bracket of
functions on $\Pi T^*M$ (which are multivector fields on $M$) maps
to the commutator of vector fields. In  turn, to each vector
field on any manifold we can assign a fiberwise-linear function on
the cotangent bundle so that the  commutator of vector fields maps
to the Poisson bracket of the corresponding   Hamiltonians. Hence we
have a sequence of linear maps preserving the brackets:
\begin{equation*}
    \fun(\Pi T^*M)\to \Vect(\Pi T^*M) \to \fun(T^*(\Pi T^*M))\to \fun(T^*(\Pi TM))\,,
\end{equation*}
where the last arrow is induced by the identification  $T^*(\Pi TM)
\cong T^*(\Pi T^*M)$.
We define $\a$ as  the through map. It is odd and takes
brackets to brackets.

\begin{coro} To each even $P\in \Mult(M)$ such that $[P,P]=0$ there corresponds an odd $K=K_P\in \fun(T^*(\Pi
TM))$ such that $(K,K)=0$.
\end{coro}

This odd Hamiltonian $K=K_P$ defines the higher Koszul brackets on the algebra of forms $\O(M)$ corresponding to a higher Poisson structure on $M$ defined by the multivector field $P$.

We may calculate the Hamiltonian $K$ explicitly. If $P=P(x,x^*)$,
then
\begin{equation}\label{koszul.eq.hamcoord}
K=(-1)^{\at}\der{P}{x^*_a}(x,\pi_{.})p_a+dx^a\,\der{P}{x^a}(x,\pi_{.})\,,
\end{equation}
where $\pi_{.}=(\pi_a)$ and we denote by $p_a,\pi_a$ the momenta conjugate to the coordinates $x^a,dx^a$ on $\Pi TM$, respectively. Note the linear dependence on the coordinate $dx^a$.

\begin{ex}
For the quadratic $P=\frac{1}{2}\,P^{ab}x^*_bx^*_a$ we get
\begin{equation}\label{koszul.eq.hamcoordquadr}
K=-P^{ab}\pi_{b}p_a+\frac{1}{2}\,dP^{ab}\pi_b\pi_a\,,
\end{equation}
which leads to the binary brackets
\begin{gather*}
    [x^a,x^b]_P=((K,x^a),x^b)=0\,, \quad [x^a,dx^b]_P=((K,x^a),dx^b)=-P^{ab}\,, \quad \text{and} \quad\\
    [dx^a,dx^b]_P=((K,dx^a),dx^b)= dP^{ab}\,
\end{gather*}
coinciding with~\eqref{koszul.eq.binary}. Therefore in this case our construction reproduces the classical
Koszul bracket.
\end{ex}

In general, the odd Hamiltonian $K=K_P$ defines a sequence of odd
$n$-ary brackets on $\O(M)$,
\begin{equation}\label{koszul.eq.higher}
    [\o_1,\ldots,\o_n]_P=(\ldots(K,\o_1),\ldots,\o_n)|_{\Pi TM}\,, \quad  n=0,1,2,\ldots \ ,
\end{equation}
which makes it a particular case of a
`homotopy Schouten algebra'. (It is an $L_{\infty}$-algebra such that  each
bracket is a multiderivation of the associative multiplication.) It is instructive to have a look at the $(k+l)$-bracket of $k$ functions $f_1,\ldots,f_k$ and $l$ differentials $df_{k+1},\ldots, df_{k+l}$. Either from~\eqref{koszul.eq.hamcoord} or directly from the construction in Theorem~\ref{koszul.thm.constr} we obtain the following formulas:
\begin{align}
    [f]_P&=\{f\}_P \quad \text{and} \quad [f_1,\ldots,f_k]=0 \ \text{for $k\geq 2$}\,,  \label{koszul.eq.higherfunc}\\
    [f_1,df_2,\ldots,df_n]_P&=(-1)^{\e}\,\{f_1,f_2,\ldots,f_n\}\,,  \label{koszul.eq.higherfuncdif}\\
    [df_1,\ldots,df_n]_P&=(-1)^{\e+1}\,d\{f_1,\ldots,f_n\}\,,  \label{koszul.eq.higherdifdif}
\end{align}
where $\e=(n-1)\ft_1+(n-2)\ft_2+\ldots+\ft_{n-1}+n$. (We also have
$[\varnothing]_P=\{\varnothing\}_P$ for the bracket without
arguments.) From  here we see that our constructions yield precisely
an \emph{$L_{\infty}$-algebroid} structure on the cotangent bundle
$T^*M$. Such a structure on a vector bundle $E$ consists of a
sequence of higher Lie brackets of sections making their space an
$L_{\infty}$-algebra and a sequence of `higher anchors' (multilinear
maps into the tangent bundle)  so that, for each $n$, the $n$-anchor
appears in the Leibniz formula for the $n$-bracket. Since for the
cotangent bundle, the differentials of functions span the space of
sections over functions,  it is sufficient to know the brackets as
well as the action of the anchors just for differentials, as the
rest can be recovered by the Leibniz rule. Therefore, taken together
with~\eqref{koszul.eq.higherfunc},
formulas~\eqref{koszul.eq.higherfuncdif} define the anchors and
formulas~\eqref{koszul.eq.higherdifdif} and
\eqref{koszul.eq.higherfuncdif}, the brackets of sections for
$T^*M$. This extends the classical construction for ordinary Poisson
manifolds, see~\cite{mackenzie:book2005}. Note finally that an
$L_{\infty}$-algebroid structure on an arbitrary $E$ is defined by a
homological vector field on the total space $\Pi E$ (for ordinary
Lie algebroids this field has to be homogeneous of degree $+1$).
What is this field in our case? One can immediately see that it is
just the odd Hamiltonian vector field $X_P\in \Vect(\Pi T^*M)$
corresponding to the function $P\in \fun(\Pi T^*M)$. This gives an
alternative proof of Theorem~\ref{koszul.thm.constr}. The higher
Koszul brackets on $\O(M)$ appear simply as the extension  of the
Lie brackets in this $L_{\infty}$-algebroid to the algebra $\O(M)$
as multiderivations, in a complete analogy with the classical case.

A question remains about the arrow $\f^*_P\co \O(M)\to \Mult(M)$.
Since there is only one non-zero bracket on $\Mult(M)$ and a whole
sequence of brackets on $\O(M)$, it cannot just map brackets to
brackets as in the classical case. A hope is that it extends to an
$L_{\infty}$-morphism. (This will be studied elsewhere.)

\section{Discussion}

Instead of a Poisson manifold (with an even Poisson structure), one
may consider an odd Poisson manifold. There is an analog of
diagram~\eqref{intro.eq.diagram} and of the Koszul
brackets~\cite{tv:laplace2}. Considerations of this paper can be
extended to this case as well yielding  `homotopy odd symplectic
structures' and higher Koszul brackets for higher Schouten
structures. This corresponds to a map $T^*M\to \Pi TM$. Note that a
map $T^*M\to TM$ is what is used in classical mechanics when passing
from the Lagrangian to the Hamiltonian picture and back; it makes
sense to study a map $\Pi T^*M\to TM$. Finally, one may wish to
replace Poisson manifolds by Lie bialgebroids and their analogs. No
doubt that the constructions of this paper can be carried over to
them as well.


\begin{theacknowledgments}
It is a pleasure to thank the organizers of the annual Workshops  on
Geometric Methods in Physics in Bia{\l}owie\.{z}a where this work
was first reported and particularly Prof. Anatol~Odzijewicz for the
hospitality and the exceptionally inspiring atmosphere at the
meetings. Most cordial thanks are due to Prof. James Stasheff for
numerous remarks on the first version of the text and his help in
improving the exposition.
\end{theacknowledgments}





\end{document}